\documentstyle[prb,aps,multicol,psfig]{revtex}

\begin{document}

\title{Environment Dependent Interatomic Potential for Bulk Silicon}

\author{Martin Z. Bazant$^\dagger$, Efthimios Kaxiras$^\dagger$
and J. F. Justo$^\ddagger$}

\address{$^\dagger$ Department of Physics, 
Harvard University, Cambridge, MA 02138}
\address{$^\ddagger$ Department of Nuclear Engineering, Massachusetts
Institute of Technology, Cambridge, MA 02139}
           
\vspace{0.2cm}
            
\date{April 16, 1997}

\maketitle
\begin{abstract}

	We use recent theoretical advances to develop a new functional
form for interatomic forces in bulk silicon.  The theoretical results
underlying the model include a novel analysis of elastic properties
for the diamond and graphitic structures and inversions of {\it ab
initio} cohesive energy curves.  The interaction model includes
two-body and three-body terms which depend on the local atomic
environment through an effective coordination number.  This
formulation is able to capture successfully: (i) the energetics and
elastic properties of the ground state diamond lattice; (ii) the
covalent re-hybridization of undercoordinated atoms; (iii) and a
smooth transition to metallic bonding for overcoordinated
atoms. Because the essential features of chemical bonding in the bulk
are built into the functional form, this model promises to be useful
for describing interatomic forces in silicon bulk phases and defects.
Although this functional form is remarkably realistic by usual
standards, it contains a small number of fitting parameters and
requires computational effort comparable to the most efficient
existing models.  In a companion paper, a complete parameterization of
the model is given, and excellent performance for condensed phases and
bulk defects is demonstrated.

\end{abstract}          
  

%\narrowtext 

\begin{multicols}{2}

\section{Introduction}
 
	The study of materials properties is increasingly relying on a
microscopic description of the underlying atomic structure and
dynamics.  While many of the key features can be described by a small
number of atoms that are actively participating in a physical process,
many problems of interest require of order 10$^3$--10$^6$ or even
higher number of atoms and time scales of 10-100 $ps$ for a proper
description.  {\it Ab initio} methods based on density functional
theory \cite{dft} and the local density approximation (DFT/LDA) have
been intensively and successfully used to provide a microscopic
description of simple structures \cite{payne}.  For more complex
cases, including for instance disordered or stepped surfaces,
dislocations, grain boundaries, crystal growth and the
amorphous--to--crystal transition, a large number of atoms is
required, making an {\it ab initio} description untenable.  A possible
alternative for these cases might be empirical interatomic potentials
which are computationally much less expensive.  The difficulty in
employing empirical potentials is their unproven ability to capture
the physics of structures far from the fitting data used to construct
them.  Developing reliable empirical potentials remains an issue of
great interest and possibly of great rewards.

	Silicon is a test case for the development of empirical
potentials for covalent materials.  Its great technological
importance, the vast amount of relevant experimental and theoretical
studies available, and its intrinsic interest as the representative
covalent material make it an ideal candidate for exploring to what
extent the empirical potential approach can be exploited. In recent
years, more than 30 empirical potentials for silicon have been
developed and applied to a number of different systems, and more
recently compared to each other \cite{bala,cook}.  They differ in
degree of sophistication, functional form, fitting strategy and range
of interaction, and each can accurately model various special atomic
configurations. Surfaces and small clusters are the most difficult to
handle \cite{bala,kax}, but even bulk material (crystalline and
amorphous phases, solid defects and the liquid phase) has resisted a
transferable description by a single potential.  Realistic simulations
of important bulk phenomena such as plastic deformation, diffusion and
crystallization are still problematic.

	In this article, we derive a general model for the functional
form of interatomic forces in bulk tetrahedral semiconductors. This
functional form is applied to the prototypical case of silicon in a
companion article \cite{justo}. The development of the model is
organized as follows: In section \ref{sec:review}, we briefly review
existing potentials and approximations of quantum models for silicon
and extract important conclusions about the desirable features of a
successful interatomic potential.  Recent theoretical advances used in
deriving our model from {\it ab initio} total energy data are outlined
in section \ref{sec:theory}.  A functional form that incorporates the
theoretical results using a minimal number of fitting parameters is
presented and discussed in section \ref{sec:form}. Finally, section
\ref{sec:concl} contains some concluding remarks.

\section{Review of Empirical Potentials and Approximations}
\label{sec:review}

\subsection{Empirical Potentials}

	The usual approach for deriving empirical potentials is to
guess a functional form, motivated by physical intuition, and then to
adjust parameters to fit {\it ab initio} total energy data for various
atomic structures. A covalent material presents a difficult challenge
because complex quantum-mechanical effects such as chemical bond
formation and rupture, hybridization, metalization, charge transfer
and bond bending must be described by an effective interaction between
atoms in which the electronic degrees of freedom have somehow been
\lq\lq integrated out'' \cite{carlsson}.  In the case of Si, the
abundance of potentials in the literature illustrates the difficulty
of the problem and lack of specific theoretical guidance. In spite of the
wide range of functional forms and fitting strategies, all proposed
models possess comparable (and insufficient) overall accuracy
\cite{bala}.  It has proven almost impossible to attribute the
successes or failures of a potential to specific features of its
functional form. Nevertheless, much can be learned from past
experience, and it is clear that a well-chosen functional form is more
useful than elaborate fitting strategies.

	To appreciate this point we compare and contrast some
representative potentials for silicon. The pioneering potential of
Stillinger and Weber (SW) has only eight parameters and was fitted to
a few experimental properties of solid cubic diamond and liquid
silicon \cite{sw}.  The model takes the form of a third order cluster
potential \cite{carlsson} in which the total energy of an atomic
configuration $\{\vec{R}_{ij}\}$ is expressed as a linear combination
of two- and three-body terms,
\begin{equation}
E = \sum_{ij} V_2(R_{ij}) + \sum_{ijk} V_3(\vec{R}_{ij},\vec{R}_{ik}),
\label{eq:clpot}
\end{equation}
where $\vec{R}_{ij} = \vec{R}_j - \vec{R}_i$, $R_{ij} =
|\vec{R}_{ij}|$ and we use the convention that multiple summation is
over all permutations of distinct indices.  The range of the SW
potential is just short of the second neighbor distance in the
equilibrium DC lattice, so the pair interaction $V_2(r)$ has a deep
well at the first neighbor distance to represent the restoring force
against stretching $sp^3$ hybrid covalent bonds. The three-body
interaction is expressed as a separable product of radial functions
$g(r)$ and an angular function $h(\theta)$
\begin{equation}
V_3(\vec{r}_1,\vec{r}_2) = g(r_1) g(r_2) h(l_{12}) ,
\label{eq:V3sw}
\end{equation}
where $l_{12} = \cos \theta_{12} = \vec{r}_1 \cdot\vec{r}_2 / ( r_1
r_2)$. The angular function, $h(l) = ( l + 1/3 )^2$ , has a minimum of
zero at the tetrahedral angle to represent the angular preference of
$sp^3$ bonds, and the radial function $g(r)$ decreases with distance
to reduce this effect when bonds are stretched. The SW three-body term
captures the directed nature of covalent $sp^3$ bonds in a simple way
that selects the diamond lattice over close-packed structures.
Although the various terms lose their physical significance for
distortions of the diamond lattice large enough to destroy $sp^3$
hybridization, the SW potential seems to give a reasonable description
of many states experimentally relevant, such as point defects, certain
surface structures, and the liquid and amorphous states \cite{bala}.
The SW potential continues to be a favorite choice in the literature,
due in large part to its appealing simplicity and apparent physical
content.

Another popular and innovative empirical model is the Tersoff
potential, with three versions generally called T1 \cite{t1}, T2
\cite{t2}, and T3 \cite{t3}. The original version T1 has only six
adjustable parameters, fitted to a small database of bulk polytypes.
Subsequent versions involve seven more parameters to improve elastic
properties. The Tersoff functional form is fundamentally different
from the SW form in that the strength of individual bonds is affected
by the presence of surrounding atoms.  Using Carlsson's terminology,
the Tersoff potential is a third order cluster functional
\cite{carlsson} with the cluster sums appearing in nonlinear
combinations. As suggested by theoretical arguments
\cite{ferrante,abell,pettifor}, the energy is the sum of a repulsive
pair interaction $\phi_R(r)$ and an attractive interaction $p(\zeta)
\phi_A(r)$ that depends on the local bonding environment, which is
characterized by a scalar quantity $\zeta$,
\begin{eqnarray}
E & = & \sum_{ij} \left[ \phi_R(R_{ij}) + p(\zeta_{ij}) \phi_A(R_{ij})
\right] \\ \zeta_{ij} & = & \sum_k V_3(\vec{R}_{ij}, \vec{R}_{ik}) ,
\label{eq:tersoff}
\end{eqnarray}
where the function $p(\zeta)$ represents the Pauling bond order.
The three-body interaction has the form of Eq. (\ref{eq:V3sw}) with
the important difference that the angular function, although still
positive, may not have a minimum at the tetrahedral angle. The T1, T2
and T3 angular functions are qualitatively different, possessing
minima at $180^o$, $90^o$ and $126.745^o$, respectively.  The Tersoff
format has greater theoretical justification away from the diamond
lattice than SW, but the three versions do not outperform the SW
potential overall, perhaps due to their handling of angular forces
\cite{bala}. Nevertheless, the Tersoff potential is another example of
a successful potential for bulk properties with a physically motivated
functional form and simple fitting strategy.
 
	The majority of empirical potentials fall into either the
generic SW \cite{bh,kp,chel} or Tersoff
\cite{dodson,kds,brenner,baskes,wr} formats just described, but there
are notable exceptions that provide further insight into successful
approaches for designing potentials. First, a number of potentials
possess functional forms that have either limited validity or no
physical motivation at all, suggesting that fitting without
theoretical guidance is not the optimal approach. The Valence Force
Field \cite{keating,bg} and related potentials \cite{stoneham,murrell}
(of which there are over 40 in the literature \cite{stoneham}) involve
scalar products of the vectors connecting atomic positions, an
approximation that is strictly valid only for small departures from
equilibrium. Thus, extending these models to highly distorted bonding
environments undermines their theoretical basis. The potential of
Pearson {\it et. al.} \cite{ptht}, as the authors emphasize, is not
physically motivated, but rather results from an exercise in
fitting. Their use of Lennard-Jones two-body terms and Axilrod-Teller
three-body terms, characteristic of Van der Waals forces, has no
justification for covalent materials. The potential of Mistriotis,
Flytzanis and Farantos (MFF) \cite{mff} is an interesting attempt to
include four-body interactions.  Although the importance of four-body
terms is certainly worth exploring, the inclusion of a four-body term
in a linear cluster expansion is not unique, and theoretical analysis
tends to favor nonlinear functionals \cite{carlsson,abell,pettifor}.
 
	A natural strategy to improve on the SW and Tersoff models is
to replace simple functional forms with more flexible ones and
complement them with more elaborate fitting schemes. The Bolding and
Andersen (BA) potential \cite{bolding} generalizes the Tersoff format
with over 30 adjustable parameters fit to an unusually wide range of
structures. Although it has not been thoroughly tested, the BA
potential appears to describe simultaneously bulk phases, defects,
surfaces and small clusters, a claim that no other potential can make
\cite{bala}. However, its complexity makes it difficult to interpret
physically, and since a large fitting database was used, it is unclear
whether the potential can reliably describe structures to which it was
not explicitly fit. In this vein, the spline-fitted potentials of the
Force Matching Method \cite{richards} represent the opposite extreme
of the SW and Tersoff approaches: physical motivation is bypassed in
favor of elaborate fitting.  These potentials involve complex
combinations of cubic splines, which have effectively hundreds of
adjustable parameters, and the strategy of matching forces on all
atoms in various defect structures is the most elaborate attempted
thus far. Although the method may be worth pursuing as an alternative,
it has not yet produced competitive potentials
\cite{richards_APS}. Moreover, even if a reliable potential could
result from such fitting strategies, it would make it hard to
interpret the results of atomistic simulations in terms of simple
principles of chemical bonding. Such interpretation is essential, in
our view, if physical insight is to be gained from computer
simulations.

	In spite of relentless efforts, no potential has demonstrated
a transferable description of silicon in all its forms \cite{bala}
leading us to another important conclusion: it may be too ambitious to
attempt a simultaneous fit of all of the important atomic structures
(bulk crystalline, amorphous and liquid phases, surfaces, and
clusters) since qualitatively different aspects of bonding are at work
in different types of structures.  Theory and general experience
suggest that the main ingredient needed to differentiate between
surface and bulk bonding preferences is a more sophisticated
description of the local atomic environment.  A notable example in
this respect is the innovative Thermodynamic Interatomic Force Field
(TIFF) potential of Chelikowsky {\it et. al.} \cite{tiff}, which
includes a quantity called the \lq\lq dangling bond vector'' that is a
weighted average of the vectors pointing to the neighbors of an
atom. For symmetric configurations characteristic of the ideal (or
slightly distorted) bulk material, the dangling bond vector vanishes
(or is exceedingly small). Conversely, a nonzero value of the dangling
bond vector indicates an asymmetric distribution of neighbors.  While
the TIFF dangling bond vector description appears to be very useful
for undercoordinated structures like surfaces and small clusters, in
this work we restrict ourselves to bulk material and thus use a
simpler, scalar environment description.  Our goal is to obtain the
best possible description of condensed phases and defects with a
simple, theoretically justified functional form.

\subsection{Approximation of Quantum Models}

	An alternative to fitting guessed functional forms is to
derive potentials by systematic approximation of quantum-mechanical
models. So far, this approach has failed to produce superior
potentials, but important connections between electronic structure and
effective interatomic potentials have been revealed.  Although
attempts are being made to directly approximate Density Functional
Theory \cite{mccarley_APS}, the most useful contributions involve
approximating various Tight Binding (TB) models, which can themselves
be derived as approximations of first principles theories
\cite{harris}. These methods are based on low order moment
approximations of the TB local density of states (LDOS), which is used
to express the average band energy as the sum of occupied bonding
states \cite{carlsson,pettifor,bop,cfm,carls,ca,harr}.  Pettifor has
derived a many-body potential, similar in form to the Tersoff
potential, by approximation of the TB bond order \cite{pettifor}. More
recently, an angular dependence remarkably close to the T3 angular
function has been derived for 
$\sigma$ bonding from the lowest order two-site term in
the Bond Order Potential expansion \cite{bop}, but the 
analytically derived function has a flat minimum
around $130^o$ and thus differs qualitatively with the T1 and T2
potentials. With hindsight, a simple physical principle explains 
these results:  
a $\sigma$ bond is most weakened (desaturated) by the presence of an
another atom when the resulting angle is small ($\theta < 100^o$)
because in such cases the atom lies near the bond axis, thus  
interfering with the 
$\sigma$ orbital where it is most concentrated.
Working within the
same framework of the TB LDOS, Carlsson and coworkers have derived
potentials with the Generalized Embedded Atom Method
\cite{cfm,carls,ca}.  Harrison has arrived at a similar model by
expanding the average band energy in the ratio of the width of the
bonding band to the bond-antibond splitting, the relevant small
parameter in semiconductors \cite{harr}.  These potentials resemble
the SW potential in its description of angular forces with an additive
three-body term, particularly for small distortions of the diamond
lattice.  The transition to metallic behavior in overcoordinated
structures involves interbond interactions similar to the Tersoff and
embedded atom potentials.

	Many-body potentials can be derived from quantum-mechanical
models if we restrict our attention to important small sets of
configurations. Using a basis of $sp^3$ hybrid orbitals in a TB model,
Carlsson {\it et. al.}  \cite{carlsson,cfm} have shown that a
generalization of the SW format, in which Eq. (\ref{eq:V3sw}) is
replaced by a form similar to that used by Biswas and Hamann (BH)
\cite{bh},
\begin{equation}
V_3(\vec{r}_1,\vec{r}_2) = \sum_{m=0}^2 g_m(r_1) g_m(r_2)\ l_{12}^m ,
\label{eq:V3bh}
\end{equation}
is valid in the vicinity of the equilibrium diamond lattice. In
general, the fourth moment controls the essential band gap of a
semiconductor, implying four-body interactions, but the separable,
three-body SW/BH terms are a consequence of the open topology of the
diamond lattice: the only four-atom hopping circuit between first
neighbors is the self-retracing path $i \rightarrow j \rightarrow i
\rightarrow k \rightarrow i$ \cite{carlsson}.

	We can make analogous arguments for the graphitic lattice to
draw conclusions about $sp^2$ hybrid bonds. Ignoring the weak,
long-range interaction between hexagonal planes, we can assume a TB
basis of $sp^2$ hybrid orbitals and follow Carlsson's
derivation. Because the self-retracing path is also the only first
neighbor hopping circuit in a graphitic plane, a cluster expansion
with the generic BH three-body interaction is also valid for hexagonal
configurations, with the functions in Eqs. (\ref{eq:clpot}) and
(\ref{eq:V3bh}) differing from their diamond $sp^3$ counterparts, as
described below. These calculations also suggest that a locally
valid cluster expansion should acquire strong environment dependence
for large distortions from the reference configuration
\cite{carlsson}.

	These studies provide theoretical evidence that the linear
three-body SW/BH format is appropriate near equilibrium structures,
while the nonlinear many-body Tersoff format describes general trends
across different bulk structures.  For the asymmetric configurations
found in surfaces and small clusters, these theories also suggest that
a more complicated environment dependence than Tersoff's is needed,
like the dangling bond vector of the TIFF potential
\cite{pettifor,cfm}. In conclusion, direct approximation of quantum
models can provide insight into the origins of interatomic forces, but
apparently cannot produce improved potentials. The reason may be that
the long chain of approximations connecting first principles and
empirical theories is uncontrolled, in the sense that there is no
small parameter which can provide an asymptotic bound for the
neglected terms for a wide range of configurations \cite{mu_note}.

\section{ Inversion of {\it Ab Initio} Energy Data }
\label{sec:theory}

	There are very few hard facts concerning the nature of
interatomic forces. Although there has been a proliferation of {\it ab
initio} energy and force calculations for a wide range of atomic
structures, it has proven difficult to discover any concrete
information regarding the functional form of interatomic
potentials. With the ubiquitous fitting approach, it is never clear
whether discrepancies with {\it ab initio} data result from an
incorrect functional form or simply suboptimal fitting
\cite{bala}. Thus, in addition to the practical problem of designing
potentials, it is also difficult to build a simple conceptual
framework within which to understand the complexities of chemical
bonding.  In this section, we summarize our recent efforts to extract
features of interatomic forces directly from {\it ab initio} total
energy data.  In order to investigate the global trends in bonding
across bulk structures predicted by quantum theories, we first perform
inversions of {\it ab initio} cohesive energy curves in part
\ref{sec:inversion}. By analyzing elastic properties of covalent
solids in part \ref{sec:elastic}, we then explore the cohesive forces
in certain special (high symmetry) bonding states, which can be viewed as an
inversion of {\it ab initio} energies restricted to selected important
configurations.

\subsection{Inversion of Cohesive Energy Curves}
\label{sec:inversion}

	We have recently shown that it is possible to derive effective
interatomic potentials for covalent solids directly from {\it ab
initio} data \cite{bazant,bazant_MRS}. The inversion procedure
generalizes the \lq\lq {\it ab initio} pair potential'' of Carlsson,
Gelatt and Ehrenreich \cite{cge} to many-body interactions and for
arbitrary strains 
beyond uniform volume expansion \cite{bazant_prep}. For the case
of silicon, this work provides first principles evidence in favor of
the generic bond order form of the pair interaction,
\begin{equation}
V_2(r,Z) = \phi_R(r) + p(Z) \phi_A(r) ,
\label{eq:bo}
\end{equation}
where $\phi_R(r)$ represents the short-range repulsion of atoms due to
Pauli exclusion of their electrons, $\phi_A(r)$ represents the
attractive force of bond formation, and $p(Z)$ is the bond order,
which determines the strength of the attraction as a function of the
atomic environment, measured by the coordination $Z$. The theoretical
behavior of the bond order is as follows
\cite{carlsson,abell,pettifor,carls,ca}: The ideal coordination for Si
is $Z_0 = 4$, due to its valence. As an atom becomes increasingly
overcoordinated ($ Z > Z_0$), nearby bonds become more metallic,
characterized by delocalized electrons. In terms of electronic
structure, the LDOS for overcoordinated atoms can be reasonably well
described by its scalar second moment.  It is a well established
result that the leading order behavior of the bond order is $p(Z) \sim
Z^{-1/2}$ in the second moment approximation
\cite{carlsson,pettifor,ca}. For $Z \leq Z_0$ on the other hand, a
matrix second moment treatment predicts a roughly constant bond order
(additive bond strengths) \cite{cfm}.  For small coordinations higher
moments are needed to incorporate important features of band shape
characteristic of covalent bonding, primarily the formation of a gap in
the LDOS \cite{carlsson,pettifor,cfm,carls}.  Thus,
the bond order should depart from the divergent $Z^{-1/2}$ behavior at
lower coordinations with a shoulder at the ideal coordination of $Z=Z_0$
where the transition to metallic $Z^{-1/2}$ dependence begins.

	Inversion of {\it ab initio} cohesive energy curves verifies
that trends in chemical bonding across various bulk bonding
arrangements are indeed consistent with these theoretical predictions
\cite{bazant}.  Previously, the only evidence in support of the bond
order formalism came from {\it equilibrium} bond lengths and energies
for a small set of ideal crystal structures
\cite{t1,t2,t3,abell,kds}. The inversion approach has revealed for the
first time that the bond order decomposition expressed by
Eq. (\ref{eq:bo}) is actually valid for a wide range of volumes away
from equilibrium and for a representative set of low energy crystal
structures. In addition to selecting the generic form of the pair
interaction, inversion provides a precise measure of the relative bond
orders in various local atomic configurations.  For example, the bond
order of $sp^2$ bonds involving three-fold coordinated atoms is about
5\% greater than that of four-fold coordinated $sp^3$ bonds in
silicon.

	These results have immediate implications for empirical
potentials. The main conclusion is that the generic Tersoff format is
much more realistic than the SW format for highly distorted
configurations. However, the inversion results also indicate that a
coordination-dependent pair interaction can provide a fair description
of high-symmetry crystal structures without requiring additional
many-body interactions. In particular, angular forces are only needed
to stabilize these structures under 
symmetry-breaking distortions, primarily for small coordinations.  
In order to make a quantitative connection between Tersoff's
functional form and our inverted {\it ab initio} data, 
angular contributions to the bond order must somehow be
suppressed for ideal crystal structures.

	The inversion procedure applied to explicit three-body
interactions has also led to some useful conclusions. Although it is
not always the case \cite{bazant_MRS}, inverted three-body radial
functions $g(r)$ tend to be strictly decreasing functions (like SW),
especially when an overdetermined set of input structures is used
\cite{bazant}. Inverted angular functions $h(l)$ also tend to penalize
small angles ($\theta < \pi/2$) less than most existing models, in
agreement with a comparative study of empirical potentials
\cite{bala}.  We must emphasize, however, that the results of this
section concern general trends in chemical bonding, and have little to
offer in terms of the precise nature of interatomic forces in special
atomic configurations, such as the low-energy states of hybrid
covalent bonds. To understand better these critical cases, we employ a
related inversion strategy.

\subsection{Analysis of Elastic Properties}
\label{sec:elastic}

	A useful theoretical approach to guide the development of
potentials, which has been pursued recently only by
Cowley \cite{cowley}, is to predict elastic properties
implied by generic functional forms and compare with experimental or
{\it ab initio} data.  This tool for understanding interatomic forces
dates back to the 19th century, when St. Venant showed that the
assumption of central pairwise forces supported by Cauchy and Poisson
implied a reduction in the number of independent elastic constants
from 21 to 15 \cite{love}.  The corresponding six dependencies, given
by the single equation $C_{12} = C_{44}$ if atoms are at centers of
cubic symmetry, are commonly called the Cauchy relations
\cite{love,bornhuang}. They provide a simple test for selecting which
materials can be described by a pair potential
\cite{hunt,cottrell}. Once it was realized that the Cauchy relations
are not satisfied by the experimental data for semiconductors, a
number of authors in this century, led by Born
\cite{bornrelation,born}, derived generalized Cauchy relations for
noncentral forces in the diamond structure \cite{delaunay,hunt}.
Building upon this body of work, we have recently analyzed the elastic
properties of several general classes of many-body potentials in the
diamond and graphitic crystal structures in order to gain insight into
the mechanical behavior of $sp^3$ and $sp^2$ hybrid covalent bonds,
respectively \cite{bazant_prep}. These high symmetry atomic
configurations must be accurately described by any realistic model of
interatomic forces in a tetravalent solid. Here we will only outline
results directly related to the model presented in the next section.

	{\it $sp^3$ Hybrids:} Consider one of the simplest many-body
interaction models for a tetrahedral solid, that is the generic SW
format defined in Eqs. (\ref{eq:clpot}) and (\ref{eq:V3sw}), with nearest
neighbor interactions and an angular function having a minimum of zero
at the tetrahedral angle ($h = h' = 0, h''>0$). In that case, first
considered by Harrison \cite{hunt,harr_phd}, the functional form of
the potential has only two degrees of freedom for elastic behavior,
$V_2''$ and $h''$, the curvatures of the pair interaction and of the angular
function at their respective minima \cite{cowley}.  Since cubic
symmetry allows for three independent elastic moduli, there is an
implied relation, due to Harrison,
\begin{equation}
(7 C_{11} + 2 C_{12}) C_{44} = 3 (C_{11} + 2 C_{12})(C_{11} - C_{12}).
\label{eq:harr}
\end{equation}
Using the experimental data \cite{simm} shown in Table I, the ratio of
the two sides of the Harrison relation is 1.16, indicating a
reasonable description by a simple SW model. In contrast, the
potentials with the Tersoff format, T2, T3 and Dodson (DOD)
\cite{dodson}, are far from satisfying this relation. This does not
imply rejection of the Tersoff format, because the functional form has
more than enough degrees of freedom to
exactly reproduce all the
elastic constants. However, as such, the inability of Tersoff
potentials to accurately describe elastic behavior when constrained to
fit other important properties does suggest a potential shortcoming in the
functional form.

	A more compelling reason to select the SW format over others
in the literature comes from the unrelaxed 
shear modulus $C_{44}^0$ which does not 
include relaxation of the internal degrees of 
freedom in the crystal unit cell. In the early literature on elastic forces,
unrelaxed elastic moduli were ignored, because they are not
experimentally accessible. With the advent of {\it ab initio}
calculations that predict elastic constants within a few percent of
experimental values, we can now analyze unrelaxed elastic properties as
well. Considering again the simple SW format, with its two degrees of
freedom, we have discovered another relation for the unrelaxed moduli,
\begin{equation}
4 C_{11} + 5 C_{12} = 9 C_{44}^0.
\label{eq:elastic}
\end{equation}
As shown in Table I, {\it the experimental and {\em ab initio} elastic
moduli satisfy this relation} within experimental and
computational error.  On the other hand, more general cluster
potentials and functionals, including the Tersoff format, BH and PTHT,
do not require this relation, and in fact cannot satisfy it under the
usual circumstances.  This is demonstrated in Table I and explains why
it has proven difficult to obtain good elastic properties with the
Tersoff potential \cite{T3_note}.  These results unambiguously select
the SW format with first neighbor interactions for describing small
homogeneous strains of the diamond lattice. Although imperfect,
internal relaxation with the SW format is also much better than with
other models. Combining Eqs. (\ref{eq:harr}) and (\ref{eq:elastic}),
we arrive at a relation involving all four moduli, $C_{11}$, $C_{12}$,
$C_{44}$ and $C_{44}^0$,
\begin{equation}
C_{44}^0 - C_{44} = \frac{(C_{11} + 8 C_{12})^2}{9 (7 C_{11} + 2
C_{12})},
\label{eq:c44relax}
\end{equation}
that expresses the effect of internal relaxation. If the two degrees
of freedom in the SW format are used to reproduce the experimental values
of $C_{11}$ and $C_{12}$, and thus also $C_{44}^0$ by
Eq. (\ref{eq:elastic}), then the predicted value of $C_{44}$ from
Eq. (\ref{eq:c44relax}) is 0.71 Mbar, which is only 12\% smaller than
the experimental value of 0.81 Mbar. The elastic behavior of the SW
format is quite remarkable considering it has only half of the
necessary degrees of freedom, while most other models are
overdetermined for elastic behavior. This explains the surprising fact
\cite{bala} that the SW potential gives one of the best descriptions
of elastic properties in spite of not having been fit to any elastic
constants.  We conclude that it is the superiority of the simple SW
functional form that gives the desirable properties, not a complex
fitting procedure.

	Using analytic expressions for the elastic constants it is
possible to devise a simple prescription to achieve good elastic
properties with the SW format. As a simple consequence of $h(-1/3) =
0$, the curvature of the pair potential is given by,
\begin{equation}
\phi''(r_d) = \frac{3 V_d}{4 r_d^2} (C_{11} + 2 C_{12}) .
\label{eq:phidia}
\end{equation}
The curvature of the angular function can be related to the second
shear modulus\cite{harrison},
\begin{equation}
g(r_d)^2 h''(-1/3) = \frac{9 V_d}{32} (C_{11} - C_{12}) ,
\label{eq:hdia}
\end{equation}
where $r_d$, $a_d$ and $V_d = a_d^3 / 8$ denote the equilibrium first
neighbor distance, lattice constant and atomic volume. Using the {\it
ab initio} data in Table I, the right hand sides of
Eqs. (\ref{eq:phidia}) and (\ref{eq:hdia}) evaluate to $8.1$ eV/$\AA^2$ and
$5.7$ eV, respectively. This provides a simple two-step procedure to
maintain good elastic behavior while fitting any model with the SW
format near the diamond lattice: ($i$) scale the pair interaction
$V_2(r)$ to obtain the correct bulk modulus $K = (C_{11} + 2
C_{12})/3$, and ($ii$) scale the three-body energy to set the second
shear modulus. As shown above, this will lead to perfect unrelaxed
elastic constants and only a 12\% error in $C_{44}$.

	{\it $sp^2$ Hybrids:} We have also obtained useful information
about interatomic forces due to $sp^2$ hybrid bonds from the elastic
moduli of the graphitic structure \cite{bazant_prep}. In this analysis
we neglect interplanar interactions, which are insignificant compared
to the covalent bonds within a single, hexagonal plane. Our goal is to
understand the elastic properties of $sp^2$ hybrids appearing around
three-fold coordinated atoms in a bulk environment, such as a
dislocation core or a grain boundary \cite{gra_note}.  An isolated
hexagonal plane embedded in three-dimensional space has three
independent (unrelaxed) elastic constants, $C_{11}$, $C_{12}$, and
$C_{44}^0$ with units of energy per unit area \cite{nye}. It can be
shown that $C_{44}^0 = 0$ for any three-body cluster potential or
functional, in perfect agreement with the vanishing {\it ab initio} value
\cite{bazantLDA}.  There is no relation for the remaining constants,
$C_{11}$ and $C_{12}$, implied by empirical models because each
functional form possesses at least two degrees of freedom.

	Drawing on the TB approximations described above, which
correctly predict the general form of interactions mediated by $sp^3$
hybrids, we proceed by assuming a separate three-body cluster
potential for $sp^2$ hybrids given by Eqs. (\ref{eq:clpot}) and
(\ref{eq:V3bh}). By analogy with the $sp^3$ case, we further assume
the simpler SW form of Eq. (\ref{eq:V3sw}) for the three-body
interaction, with the important difference that the angular function
has a minimum of zero at the {\it hexagonal angle} of $2\pi/3$ rather
than at the tetrahedral angle. We again restrict the interaction range to
nearest neighbors engaged in the covalent bonds that dominate
cohesion. These are not the only possible choices, but we can evaluate
their validity through analysis of elastic moduli.

	With such a functional form \cite{ik}, which differs from all
existing empirical potentials \cite{kds_note}, stability
considerations imply $C_{11} > 3 C_{12}$, which is indeed satisfied by
the {\it ab initio} values, $C_{11}=1.79$ Mbar and $C_{12}=0.51$ Mbar
\cite{bazantLDA}.
More importantly, we can relate the mechanical properties of $sp^2$
and $sp^3$ hybrids. The relative radial stiffness is given by a simple
ratio of elastic constants,
\begin{equation}
\frac{\phi''_h(r_h)}{\phi''_d(r_d)} = \frac{8 r_d^2}{9 r_h^2} 
\frac{A_h (C_{11} + C_{12})_h}{V_d (C_{11} + 2 C_{12})_d} ,
\label{eq:phiratio}
\end{equation}
where the subscript $h$ refers to the equilibrium hexagonal plane
with area per atom  $A_h = a_h^2 \sqrt{3}/4$, and $d$ refers to the
diamond lattice. 
Using the {\it ab initio} result, $r_h = 2.23$\AA, the prefactor
in Eq. (\ref{eq:phiratio}) is 0.99, so the elastic constant ratio in
parentheses provides a direct comparison of $sp^2$ and $sp^3$ radial
forces. The {\it ab initio} value of that ratio is $1.4 \pm 0.1$,
implying that $sp^2$ bonds have 40\% greater radial stiffness than
$sp^3$ bonds. The same result also follows directly from inverted pair
potentials for the graphitic and diamond structures \cite{bazant}. 

	A similar elastic analysis yields an expression for the relative
angular stiffness of $sp^2$ and $sp^3$ hybrid bonds,
\begin{equation}
\frac{h_h''(-1/2)}{h_d''(-1/3)} = \frac{256 g_d(r_d)^2}{243 g_h(r_h)^2} 
\frac{ A_h (C_{11} - 3
C_{12})_h} { V_d (C_{11} - C_{12})_d },
\label{eq:hratio}
\end{equation}
Using the {\it ab initio} data, we have the general result,
$g_h(r_h)^2 h_h''(-1/2) / g_d(r_d)^2 h_d''(-1/3) \ = \ 0.46 \pm 0.15$.
Assuming $g_d(r) \approx g_h(r)$ with each function decreasing in
accordance with inversion results \cite{bazant}, then the product of
prefactors in Eq. (\ref{eq:hratio}) is nearly unity. In that case the
ratio of elastic constants in parentheses allows us to quantify the relative
bending strength of the hybrid bonds. The {\it ab initio} value
for the ratio of $0.44 \pm 0.15$ indicates that the angular stiffness
of $sp^2$ bonds is smaller than that of $sp^3$ bonds by about a factor
of two, in spite of the greater radial stiffness of $sp^2$ bonds. Our
conclusion for the relative bending strength of $sp^2$ and $sp^3$
hybrids would be reversed only if $g_g(r_g)$ were smaller than
$g_d(r_d)$ by at least a factor of two, which seems unlikely in light
of the bond orders.

	 Elastic constant analysis suggests that a hybrid covalent
bond is well represented by a separable, first-neighbor, three-body
cluster potential whose angular function has a minimum of zero at the
appropriate angle.  This may seem to contradict the ample evidence we
have cited in favor of the Tersoff format for large distortions of the
diamond lattice, particularly those involving changes in
coordination. These findings are consistent, however, in light of
Carlsson's argument that cluster potentials like SW can accurately fit
narrow ranges of configurations while cluster functionals like
Tersoff's provide a less accurate but physically acceptable fit to a
much broader set of configurations \cite{carlsson_note}.

	This body of results forms a reliable foundation upon which to
build empirical potentials for bulk tetravalent solids. In general, we
conclude that the functional form of atomic interactions should reduce
exactly to appropriate cluster potentials in special bonding
geometries, with environment dependence that interpolates smoothly
between these special cases and captures general trends.  We shall
refer to this theoretically motivated functional form as the
Environment Dependent Interatomic Potential (EDIP) for Bulk Si.

\section{Functional Form}
\label{sec:form}

	Although reasonable interaction potentials can be derived
using the analytic methods of the previous section, such inversion
schemes become most powerful when used as theoretical guidance for
fitting. The reason is that inversion necessarily involves a
restricted set of {\it ab initio} data.  Although the input data can
be perfectly reproduced (unless it is overdetermined), it is desirable
to allow an imperfect description of the inversion data in order to
achieve a better overall fit of a wider {\it ab initio} database that
includes low symmetry defect structures.  Thus, our approach is to
incorporate the theoretically derived features of the previous section
directly into our functional form, and then to fit the potential to a
carefully chosen {\it ab initio} database with a minimal number of
parameters.  In this way, we can systematically derive a reliable
potential for bulk properties while keeping the functional form simple
enough to allow for efficient computation of forces as well as
intuitive understanding of chemical bonding in covalent solids.

\subsection{Scalar Environment Description}

	The simplest description of the local environment of an atom
is the number of nearest neighbors, determined by an effective
coordination number $Z_i$ for atom $i$,
\begin{equation}
Z_i = \sum_{m \neq i} f(R_{im})
\end{equation}
where $f(R_{im})$ is a cutoff function that measures the contribution
of neighbor $m$ to the coordination of $i$ in terms of the bond length
$R_{im}$. The special $sp^2$ and $sp^3$ bonding geometries can be
uniquely specified by their coordinations due to their high
symmetry. Since environment dependence is not needed in those cases,
it is natural to take the coordination number to be a constant, except
when large distortions from equilibrium occur.  Moreover, covalent
bonds tend to involve only first neighbors, as indicated by {\it ab
initio} charge density calculations of open structures like the
diamond lattice \cite{kaxboyer}. Thus, we choose the neighbor function
to be exactly unity for typical covalent bond lengths, $r<c$, with a
gentle drop to zero above a cutoff $b$ that excludes second neighbors,
\begin{equation}
f(r) = \left\{ \begin{array}{ll} 1 & \mbox{if $r < c$} \\
               \exp\left(\frac{\alpha}{1 - x^{-3}}\right) & \mbox{if
               $c < r < b$} \\ 0 & \mbox{if $r > b$} \end{array}
               \right.  .
\end{equation}
where $x = (r- c)/(b - c)$.  This particular choice of cutoff
function is appealing because it has two continuous derivatives at the
inner cutoff $c$, and is perfectly smooth at the outer cutoff $b$. The
cutoffs $b$ and $c$ are restricted to lie between first and second
neighbors of both the hexagonal plane and diamond lattice in
equilibrium, so that their coordinations are 3 and 4, respectively.

	Our scalar description of the atomic environment is similar to
Tersoff's, but there are notable differences. First, the perspective
is that of the atom rather than the bond: With our potential, the
preferences for special bond angles, bond strengths and angular forces
are the same for all bonds involving a particular atom. This is in
contrast to the Tersoff format \cite{t1,t2,t3,dodson,bolding} in which
a mixed bond--atom perspective is adopted: the contribution of atom
$i$ to the strength of bond $(ij)$ is affected by the ``interference''
of other bonds $(ik)$ involving atom $i$. This model provides an
intuitive explanation for trends in chemical reaction paths of
molecules \cite{brenner2} and allows for both covalent and metallic
bonds to be centered at the same atom, as observed, for example, in
{\it ab initio} charge densities for the BCT5 lattice \cite{kaxboyer},
which lies between the covalent diamond lattice and the metallic
$\beta$-tin lattice.  However, the analysis of elastic properties
discussed earlier favors the present approach for environment
dependence near the diamond lattice.  Another important difference
between our model and Tersoff's is the separation of angular
dependence from the bond order.  As we shall see, this allows us to
control independently the preferences for bond strengths, bond angles,
and angular forces in a way that the Tersoff potential cannot. By
keeping the bond order simple, we can also directly use the important
theoretical results that motivated the Tersoff potential in the first
place.

\subsection{Coordination-Dependent Chemical Bonding}

	Our potential consists of coordination-dependent two- and
three-body interactions corresponding to the defining features of
covalent materials: {\it pair bonding and angular forces}. The energy
of a configuration $\{\vec{R}_i\}$ is a sum over single-atom energies,
$E = \sum_i E_i$, each expressed as a sum of pair and three-body
interactions
\begin{equation}
E_i = \sum_j V_2(R_{ij}, Z_i) + \sum_{jk}
 V_3(\vec{R}_{ij},\vec{R}_{ik}, Z_i) ,
\end{equation}
depending on the coordination $Z_i$ of the central atom.  The pair
functional $V_2(R_{ij}, Z_i)$ represents the strength of bond $(ij)$,
while the three-body functional $V_3(\vec{R}_{ij},\vec{R}_{ik}, Z_i)$
represents preferences for special bond angles, due to hybridization,
as well as the angular forces that resist bending away from those
angles. From our atomic perspective, the pair interaction is broken
into a sum of contributions from each atom, and similarly the
three-body interaction is broken into a sum over the three angles in
each triangle of atoms. Note that due to the environment dependence,
the contributions to the bond strength from each pair of atoms are not
symmetric in general, $V_2(R_{ij}, Z_i) \neq V_2(R_{ji}, Z_j)$.

	{\it Pair Bonding:} We adopt the well-established bond order
format of Eq. (\ref{eq:bo}) for the pair interaction.  Drawing on the
popularity of the SW potential, we use those functional forms for the
attractive and repulsive interactions,
\begin{equation}
V_2(r,Z) = A \left[ \left(\frac{B}{r}\right)^\rho - p(Z) \right] \exp
\left( \frac{\sigma}{r-a} \right) ,
\end{equation}
which go to zero at the cutoff $r=a$ with all derivatives
continuous. This choice can reproduce the shapes of inverted pair
potentials for silicon \cite{bazant}.  Because we have constructed
$Z$, and hence $p(Z)$, to be constant near the diamond lattice, our
pair interaction reduces exactly to the SW form for configurations
near equilibrium, thus allowing us to obtain excellent elastic
properties as explained above.  Making this choice of repulsive term
with the parameters obtained by fitting to defect structures
\cite{justo}, we can follow the procedure of Bazant and Kaxiras
\cite{bazant} to 
extract the implied bond order $p(Z)$ from {\it ab initio} cohesive
energy curves for the following crystal structures (with coordinations
given in parentheses): graphitic (3), diamond (4), BC-8 (4), BCT-5
(5), $\beta$-tin (6), SC (6) and BCC (8).  These structures span the
full range from three and four-fold coordinated covalent bonding in
$sp^2$ and $sp^3$ arrangements, to overcoordinated atoms in metallic
phases.  The inverted {\it ab initio} bond order versus coordination
is shown in Fig. 1, along with two additional data points.
Since we have only first neighbor interactions in the diamond lattice,
we can obtain another bond order for three-fold coordination from the
{\it ab initio} formation energy (3.3 eV) for an unrelaxed vacancy.
An additional data point for unit coordination comes from the
experimental binding energy (3.24 eV) and bond length (2.246 \AA) of
the Si$_2$ molecule \cite{huber}. 

	The bond order data has a clear shoulder at $Z=Z_0=4$ where
the predicted transition from covalent to metallic bonding occurs.
For overcoordinated atoms with $Z > Z_0$, the bond order approaches
its rough asymptotic behavior, $p \propto Z^{-1/2}$, characteristic
of metallic band structure.  For coordinations $Z \leq Z_0$, the bond
order departs from the $Z^{-1/2}$ divergence, due to the formation
of a band gap in the LDOS associated with covalent bonds.  A natural
choice to capture this shape is a Gaussian, $ p(Z) = \mbox{e}^{-\beta
Z^2}$ .  In Fig. 1, we see that the bond order function we
obtain from fitting \cite{justo} is fairly close to the inversion
data. It is intentionally somewhat too large for coordinations 5--8 to
compensate for the small, but nonvanishing many-body energy for those
structures, as described below.  The collapse of the attractive
functions $\phi_A(r) = (V_2(r,Z) - V_A(r))/p(Z)$ with this choice of
bond order shown in Fig. 2 is reasonably good, thus
justifying the bond order formalism across a wide range of volumes.
Our potential is the first to have a bond order in such close
agreement with theory, which is a direct result of our novel treatment
of angular forces.

	{\it Angular Terms:} In a thorough comparative study of Si
potentials, Balamane {\it et. al.} attribute the limitations of
empirical models to the inadequate description of angular forces
\cite{bala}. Our potential contains a number of innovations in
handling angular forces, leading to a significant improvement over
existing models in reproducing {\it ab initio} data.  Analysis of
elastic properties shows that, at least near equilibrium, the
three-body functional should be expressed as a single, separable
product of a radial function $g(r)$ for both bonds and an angular
function $h(\theta,Z)$,
\begin{equation}
V_3(\vec{R}_{ij},\vec{R}_{ik}, Z_i) = g(R_{ij}) g(R_{ik})
h(l_{ijk},Z_i) .
\end{equation}
Although the radial functions could vary with coordination, in the
interest of simplicity we have focused on the angular function as the
most important source of coordination dependence.  Inversion of {\it
ab initio} cohesive energy curves \cite{bazant} suggests that a
consistent choice for the radial functions is the monotonic SW form,
\begin{equation}
g(r) = \exp \left( \frac{\gamma}{r-b}\right) ,
\end{equation}
which also goes to zero smoothly at a cutoff distance $b$, a value
different from the two body cutoff $a$.  Having separate cutoffs for
two and three-body interactions is reasonable because they describe
fundamentally different features of bonding. Although the pair
interaction might extend considerably beyond the equilibrium first
neighbor distance, the angular forces should not be allowed to extend
beyond first neighbors, if they are to be interpreted as representing
the resistance to bending of covalent bonds.

	Much of the new physics contained in our potential comes from
the angular function $h(l,Z)$. Theoretical considerations lead us to
postulate the following general form:
\begin{equation}
h(l,Z) = H\left( \frac{ l + \tau(Z) }{w(Z)} \right) ,
\end{equation}
where $H(x)$, $w(Z)$ and $\tau(Z)$ are generic functions whose
essential properties we now describe. The overall shape of the angular
function is given by $H(x)$, a nonnegative \cite{ca,carlsson} function
with a quadratic minimum of zero at the origin, $H(0) = H'(0) =0$ and
$H''(0)>0$. The function $H(x)$ should also become flat away from
the minimum well at the origin, resulting in zero angular force for
large distortions away from equilibrium.  This
feature, which is absent in most potentials including SW, is essential
for the angular term to have physical meaning far from
equilibrium. When covalent bonds are strongly bent from their
equilibrium angle they are weakened and replaced by new electronic
states. Thus, for large angular distortions it is not possible to
define a restoring force that drives atoms back towards an equilibrium
bond angle.  These properties of $H(x)$ can be satisfied by the
following choice,
\begin{equation}
H(x) = \lambda \left( 1 - \mbox{e}^{-x^2} \right) ,
\end{equation}
which is similar in shape to the MFF angular function
\cite{mff}. However, our angular dependence is considerably more
sophisticated than MFF due to its environment dependence.

	Motivated by theory, we choose the function $\tau(Z)$ to
control the coordination-dependent minimum of the angular function,
$l_0(Z) = \cos({\theta_0(Z)}) = - \tau(Z)$, with the following form
\cite{ik,kds_note},
\begin{equation}
\tau (Z) = u_1 + u_2 (u_3 e^{-u_4 Z} - e^{-2 u_4 Z}) .
\end{equation}
The parameters, $ u_1 = -0.165799$ , $u_2 = 32.557$, $u_3 = 0.286198$,
and $u_4 = 0.66$, were chosen to make the preferred angle $\theta_0(Z) =
\cos^{-1}[-\tau(Z)]$ interpolate smoothly
between several theoretically motivated values, as shown in
Fig. 3: We have already argued that $\tau(4) = 1/3$ and
$\tau(3) = 1/2$ (so that $sp^3$ and $sp^2$ bonding correspond to the
diamond and graphitic structures respectively), which determines two
of the four parameters in $\tau(z)$.  The remaining two parameters
were selected so that $\tau(2) = \tau(6) = 0$ or $\theta_0(2) =
\theta_0(6) = \pi/2$. For two-fold coordination, this choice
reproduces the preference for bonding along two orthogonal $p$-states
with the low energy, nonbonding $s$ state fully occupied. For six-fold
coordination, the 
choice $\theta_0(6) = \pi/2$ also reflects the $p$ character of the
bonds. However, structures with $Z=6$ like SC and $\beta$-tin are
metallic, with delocalized electrons that tend to invalidate the
concept of bond-bending underlying the angular function, a crucial
point we shall address shortly. The vanishing many-body energies for
the graphitic plane and diamond structures allow fitting of the pair
interactions $V_2(r,3)$ and $V_2(r,4)$ to be guided by
Eq. (\ref{eq:phidia}), which determines $V_2''(r_d,4)$ from the bulk
modulus, and Eq. (\ref{eq:phiratio}), which requires $V_2''(r_h,3) /
V_2''(r_d,4) \approx 1.4$. Moreover, the shifting of the minimum of the
angular function in our model incorporates coordination-dependent
hybridization in a way that other potentials cannot.

	Through the function $w(Z)$, our angular function has another
novel coordination dependence to represent the covalent to metallic
transition. The width of the minimum $w(Z)$ is broadened with
increasing coordination, thus reducing the angular stiffness of the
bonds as they become more metallic. Similarly, as coordination is
decreased from 4 to 3, the width of the minimum is increased to
reproduce the smaller angular stiffness of $sp^2$ bonds compared to
that of $sp^3$ bonds.  Thus, the function $w(Z)$ should have a minimum
at $Z_0=4$ and diverge with increasing $Z$.  Fitting of the model can
be guided by Eq. (\ref{eq:hdia}), which determines $w(4)$ from the
second shear modulus, and by Eq.(\ref{eq:hratio}), which requires
$w(3) / w(4) \approx \sqrt{2}$.  The softening of the angular function
is important because it allows the decrease in cohesive energy per
atom concomitant with overcoordination to be modeled by a weakening of
pair interactions. In contrast, cluster potentials like SW penalize
overcoordinated structures with increased three-body energy that
overcomes the decrease in pair bonding energy.  This is an unphysical
feature, since overcoordinated structures do not even have covalent
bonds, and the many-body energy cannot be viewed as a consequence of
stretching $sp^3$ bonds far from the tetrahedral geometry. In this
sense, the reasonably good description of liquid Si (a metal with
about 6 neighbors per atom) with the SW potential appears to be
fortuitous.

	The coordination dependence of our angular function makes it
possible for the first time to reproduce the well-known behavior of
the bond order. The reason is that the contribution of the three-body
functional to the total energy is suppressed for ideal crystals and
overcoordinated structures. The shifting of the minimum makes the
three-body energy vanish identically for $sp^2$ and $sp^3$ hybrids,
and the variable width greatly reduces the three-body energy in
metallic structures. With the three-body energy suppressed, we can use
our knowledge of the bond order for the graphitic, diamond,
$\beta$-tin and other lattices from inversion of cohesive energy
curves to capture the energetics of these structures in the pair
interaction, as described above.  Several other potentials have tried
to incorporate the bond order predicted from theory, but the
uncontrolled many-body energy makes it impossible to connect directly
with theory. Our treatment of angular forces is intuitively appealing
because the forces primarily model the bending of covalent bonds, with
the control of global energetics left to the pair interactions.

	Although our model contains a complicated environment
dependence, forces can still be evaluated with computational speed
comparable to much simpler existing potentials. The coordination
dependence introduces an extra loop into each force calculation. For
the three-body functional, this introduces a fourth nested loop over
atoms $m$ outside each triplet $(ijk)$ that contribute to coordination
of atoms $i$, which would make force evaluation much slower than the
typical three-body cluster expansions used in most other
potentials. However, our choice of $f(r)$ greatly reduces the
frequency of four body computations because nonzero forces result if
the fourth atom lies in the range of being a partial neighbor, $c <
r_{im} < b$, which happens only for a small number of neighbors in
most cases. If coordinations stay relatively constant during a
simulation, as in a low temperature solid, the four-body force
computation is insignificant. Indeed, we have found that force
evaluation with our model can be almost as fast as with the SW
potential \cite{justo}, which is an advantage of our model over others
of comparable sophistication.

\section{Conclusion}
\label{sec:concl}

	In summary, we have used recent theoretical innovations to
arrive at a functional form that describes the dependence of chemical
bonding on the local coordination number. Bond order, hybridization,
metalization and angular stiffness are all described in qualitative
agreement with theory. Consistent with our motivation, we have kept
the form as simple as possible, reproducing the essential physics with
little more complexity than existing potentials. The fitted
implementation of the model described in the companion paper
\cite{justo} involves only 13 adjustable parameters. Using the results
of the present article, we provide theoretical estimates of almost half of
the parameters, thus greatly narrowing the region of parameter space
to be explored during fitting. The remaining parameters are chosen to
fit important bulk defect structures.

	Considering the theory behind our model, we can anticipate its
range of applicability. We have shown that the structure and
energetics of the diamond lattice can be almost perfectly
reproduced. Because small distortions of $sp^3$ hybrids are accurately
modeled, we would also expect a good description of the amorphous
phase. Defect structures involving $sp^2$ hybridization should also be
well described. In general, the model should perform best whenever the
coordination number can adequately specify the local atomic
environment. This certainly includes $sp^2$ and $sp^3$ hybridization
and some metallic states, but might also include more general
situations in which atoms are more or less symmetrically distributed,
like the liquid and amorphous phases and reconstructed dislocation
cores and grain boundaries. The theory behind the model begins to
break down for noninteger coordinations, since our effective
coordination number is a way of smoothly interpolating between
well-understood local structures. More seriously, no attempt is made
to handle asymmetric distributions of neighbors, which are abundant in
surfaces and small clusters.  Theory suggests that our model may be
fitted to provide a good description of condensed phases and defects
in bulk tetrahedral semiconductors, such as Si, Ge and with minor
extensions perhaps alloys such as SiGe, that can be understood in
terms of simple principles of covalent bonding.

\acknowledgments Partial support was provided to MZB by a
Computational Science Graduate Fellowship from the Office of
Scientific Computing of the U.S. Department of Energy and to JFJ by
the  Brazilian Agency CNPq and
by the MRSEC Program of the National Science Foundation under award
number DMR 94-00334.

\end{multicols}

\vfill
\eject

\begin{center}
\begin{table}
\caption{ Comparison of elastic constants (in units of Mbar) for
diamond cubic silicon computed from empirical models with experimental
or {\it ab initio} (LDA) values.
The values for experiment (EXPT) are from Simmons and Wang
\cite{simm}, for tight-binding (TB) from Bernstein and Kaxiras
\cite{noam} and for the empirical potentials Biswas-Haman (BH), Tersoff
(T2, T3), Dodson (DOD) and Pearson-Takai-Halicioglu-Tiller (PTHT) from
Balamane \cite{bala}.  The Stillinger-Weber (SW) values were
calculated with the analytic formulae of Cowley \cite{cowley} and
scaled to set the binding energy to 4.63 eV \cite{bala}.  In the lower
half of the table, we test the elastic constant relations discussed in
the text by calculating the ratios $\alpha_H \equiv (7 C_{11} + 2
C_{12}) C_{44}/ 3 (C_{11} + 2 C_{12})(C_{11}-C_{12})$ and $\alpha_B
\equiv (4 C_{11} + 5 C_{12})/9 C_{44}^0$.
}
\begin{tabular}{cccccccccc}
         & EXPT & LDA & SW    & BH    & T2    & T3    & DOD   & PTHT  & TB \\
\hline
$C_{11}$ & 1.67 & & 1.617 & 2.042 & 1.217 & 1.425 & 1.206 & 2.969 & 1.45 \\
$C_{12}$ & 0.65 & & 0.816 & 1.517 & 0.858 & 0.754 & 0.722 & 2.697 & 0.845 \\
$C_{44}$ & 0.81 & & 0.603 & 0.451 & 0.103 & 0.690 & 0.659 & 0.446 & 0.534 \\
$C_{44}^0$ & & 1.11 & 1.172 & 1.049 & 0.923 & 1.188 & 3.475 & 2.190 & 1.35 \\
\hline
% Born relation - save for the elastic constant paper.
%  $\frac{4 C_{11} (C_{11} - C_{44})}{(C_{11} + C_{12})^2}$
%       & 1.07 & & 1.11 & 1.03 & 0.33 & 0.28 & 0.71 & 0.93 & 0.11 \\
  $\alpha_H$
        & 1.16 & & 1.00 & 0.98 & 2.99 & 2.31 & 1.69 & 1.71 & 2.80 \\
  $\alpha_B$
        & \multicolumn{2}{c}{0.99} 
	         & 1.00 & 1.67 & 1.10 & 0.89 & 0.27 & 1.29 & 0.82 \\
\end{tabular}
\end{table}
\end{center}

\begin{figure}
\begin{center}
\mbox{
\psfig{file=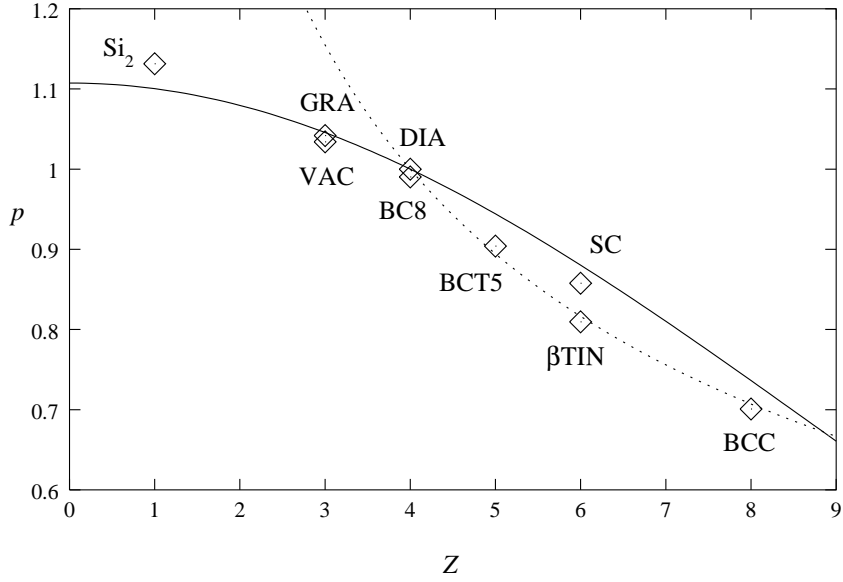,height=3in}
}
\end{center}
\caption{{\it Ab initio} values for the bond order as a function of
coordination, obtained from the inversion of cohesive energy curves
for the graphitic (GRA), cubic diamond (DIA), BC8, BCT5, SC,
$\beta$-tin and BCC bulk structures and with additional
points for the unrelaxed vacancy (VAC) and the dimer molecule
(Si$_2$), as explained in the text.  For comparison the solid
line shows the Gaussian $p(Z)$ obtained from fitting to defect
structures. The dotted line shows the $1/\sqrt{Z}$
dependence, the theoretically predicted approximate behavior for large
coordinations.}
\label{fig:p}
\end{figure}

\begin{figure}
\begin{center}
\mbox{
\psfig{file=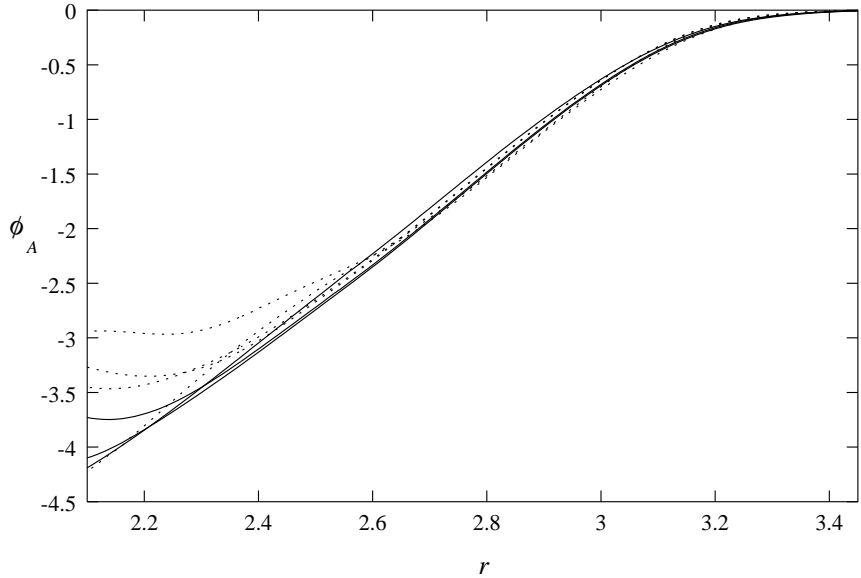,height=3in}
}
\end{center}
\caption{Attractive pair interactions from inversion of {\it ab
initio} cohesive energy curves for the structures in
Fig. 1 using the bond order and repulsive pair potential
of our model.  The solid lines are for the covalent
structures with coordinations 3 and 4, while the dotted lines are for
the overcoordinated metallic structures.  The reasonable collapse of
the attractive pair potentials indicates the validity of the bond
order functional form of the pair interaction across a wide range of
volumes and crystal structures. }
\label{fig:vaf}
\end{figure}

\begin{figure}
\begin{center}
\mbox{
\psfig{file=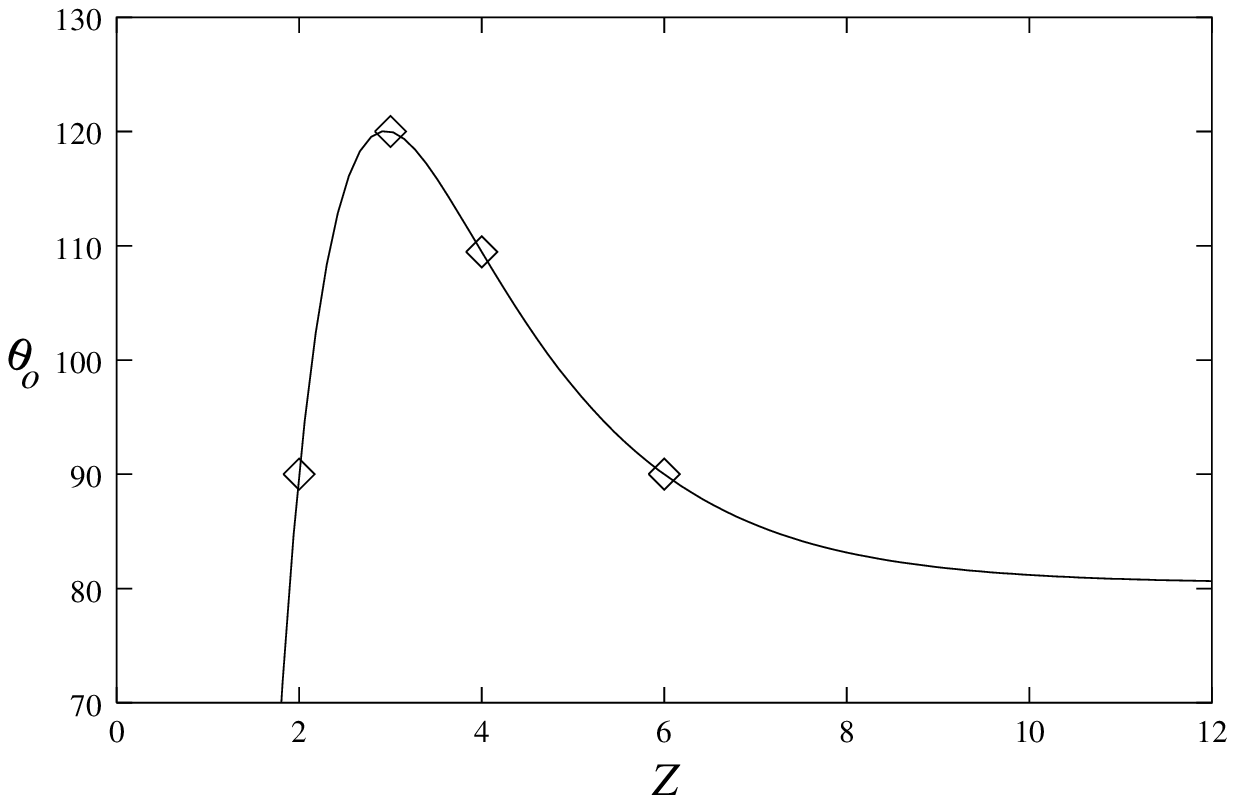,height=3in}
}
\end{center}
\caption{ The coordination dependence of the preferred bond angle 
$\theta_o(Z)$ (in degrees), which interpolates the theoretically
motivated points 
for $Z=2,3,4,6$, indicated by diamonds.
}
\end{figure}


\begin{references}
\bibitem{mu_note}   The expansion parameters in these studies are
the dimensionless ratios of high to low order moments of the
TB LDOS, which may not be small, especially for defect structures with
states in the band gap or for metallic states like the liquid,
the $\beta$-tin crystal structure, and certain surfaces. Low order moments
do capture general trends in energy, but cannot be expected to
maintain quantitative accuracy.
\bibitem{T3_note} Note that the angular function of the third Tersoff
potential\cite{t3}, which was specifically optimized for elastic
properties,  
resembles the SW angular function 
while the other versions do not.
\bibitem{gra_note}
Although the covalent bonds in a graphitic plane resemble those around
three-fold coordinated atoms in the bulk, we note that the
out-of-plane elastic properties of the graphitic structure
(e.g. $C_{44}$) are affected by nearly free $\pi$ electrons which do not
participate in the covalent $sp^2$ hybrid states
\cite{harrison}. These extra electrons, one per atom, try to stay in
perpendicular $p$-like orbitals, thus contributing to the rigidity of
the plane. Nevertheless, we expect the dominant mechanical behavior of
three-fold coordinated bulk defect atoms to be quite similar to atoms in
a graphitic plane.
\bibitem{bazantLDA} We calculated the DFT/LDA elastic moduli for an
isolated hexagonal plane using a plane wave basis with a 12 Ry cutoff
and 1296 points in the full Brillouin zone for reciprocal space
integrations. These choices guarantee sufficient accuracy. In order to
preserve periodic boundary conditions, the out-of-plane lattice
parameter was fixed at at $c = 5.5$ \AA, which is large enough to
ensure neglible interplanar forces. The parabolic regime of the
energy versus strain data (typically, strains less than 3\%) was
isolated by considering the goodness of parabolic fit, as measured by
the $\chi^2$ statistic.
\bibitem{kds_note} Khor and Das Sarma used a shifted equilibrium bond angle
within the Tersoff format, but they did not specify exactly how the
equilibrium angle should depend on the local environment in a general
configuration \cite{kds}.
\end{references}
\end{document}